\documentclass[preprint,
 amsmath,amssymb,
 aps,nofootinbib
]{revtex4}

\usepackage{graphicx}
\usepackage{dcolumn}
\usepackage{bm}

\usepackage{float}
\usepackage{amsmath,amssymb}
\usepackage{graphicx}
\usepackage{color}

\usepackage{tikz}
\usetikzlibrary{hobby}
\usepackage{hyperref}

\usepackage{graphicx}
\usepackage{amsmath,amssymb,graphicx}
\usepackage{tabularx}
\newcolumntype{C}[1]{>{\centering\arraybackslash}p{#1}}
\usepackage{hyperref}
\usepackage{comment}
\usepackage{xcolor}
\usepackage{cleveref}
\newcommand{\be}{\begin{equation}}
\newcommand{\ee}{\end{equation}}
\newcommand{\bea}{\begin{eqnarray}}
\newcommand{\eea}{\end{eqnarray}}
\newcommand{\non}{\nonumber}

\begin{document}

\title{Vacuum decay in quadratic gravity}

\author{Silvia Vicentini}

\email{silvia.vicentini@unitn.it}
\author{Massimiliano Rinaldi}%
\email{massimiliano.rinaldi@unitn.it}
\affiliation{Dipartimento di Fisica, Universit\`{a} di Trento,\\Via Sommarive 14, I-38123 Povo (TN), Italy}
\affiliation{Trento Institute for Fundamental Physics and Applications (TIFPA)-INFN,\\Via Sommarive 14, I-38123 Povo (TN), Italy}

\date{\today}

\begin{abstract} Metastable states decay at zero temperature through quantum tunneling at an exponentially small rate, which depends on the Coleman-de Luccia instanton, also known as bounce. In some theories, the bounce may not exist or its on-shell action may be ill-defined or infinite, thus hindering the vacuum decay process. In this paper, we test this possibility in modified theories of gravity interacting with a real scalar field. We consider an Einstein-Hilbert term with a non-minimally coupled scalar field and a quadratic Ricci scalar contribution. To tackle the problem we use a new analytic method, with which we prove that the scalar field on the bounce has a universal behaviour at large Euclidean radii,  almost independently of the potential. Our main result is that the quadratic Ricci scalar prevents the decay, regardless of the other terms in the action. 
\end{abstract}

\maketitle

\section{Introduction}
Coleman and De Luccia discovered long ago how to describe the decay of a metastable state using the Euclidean path integral in the semi-classical approximation\cite{Coleman:1977py,Callan:1977pt,Coleman:1980aw}. Such decay is driven by quantum fluctuations, which yield the spontaneous nucleation of a true vacuum phase in a bubbly shape. The nucleation rate is exponentially small and the exponent is proportional to the Euclidean action of the theory evaluated on a particular $O(4)$-symmetric trajectory (called Coleman-de Luccia instanton or bounce) between the tunneling point and the false vacuum. Since the equations of motion are non-linear the bounce is usually found numerically \cite{Claudson:1983et,Kusenko:1995jv,Kusenko:1996jn,Dasgupta:1996qu,Moreno:1998bq,John:1998ip,Masoumi:2016wot,Espinosa:2018hue,Espinosa:2018voj,Chigusa:2019wxb}, despite some analytical solutions exist \cite{Linde:1981zj,FerrazdeCamargo:1982sk,lee,Dutta:2011rc,Kanno:2012zf,Aravind:2014pva,Guada:2020ihz}. Analytic approximate solutions can also be found when the energy difference between the false  and the true vacuum is sufficiently small, in what is called the thin-wall approximation \cite{Coleman:1980aw,Garfinkle:1989mv,Linde:1981zj,lee,Kanno:2012zf,Eckerle:2020opg}). To demonstrate that a bounce does not exist is trickier. For instance, in the thin-wall approximation, this happens when the action has no minimum for finite non-vanishing values of the bounce radius. Most systems though cannot be studied in this approximation and the question cannot be settled. Similar results may be derived in the case of a single scalar field non-minimally coupled to gravity but only if the gravitational backreaction is small \cite{Isidori:2007vm,Salvio:2016mvj}.

Some cases of physical interest, such as modified gravity, have not been investigated much yet in the context of vacuum decay \cite{Salehian:2018yoq}. In this paper, we begin to explore systematically this class of theories and focus in particular on finding obstructions to the decay process when the false vacuum state has a flat geometry. One of our main results is that quadratic gravity terms forbid vacuum decay at zero temperature. Such term is usually required by renormalizability in quantized field theories on a gravitational (classical) background \cite{birrell,parker,Buchbinder:1992rb} and arises also as a low-energy limit of $f(R)$ gravity. Since the equations of motion, in this case, are quite involved,  we introduce a new method that allows determining the bounce at large Euclidean radii, i.e. when it approaches the false vacuum, which will be called ``asymptotic bounce" in the following. To illustrate it, we first focus on single scalar field theories on a fixed, flat background.  We then extend it to Einstein-Hilbert gravity as well as modified gravity and determine it in the following cases
\begin{itemize}
    \item  a single scalar field theory with Einstein-Hilbert gravity, a non-minimal coupling $\xi \phi^2 R$ and a quadratic term $ R^2$. 
    \item  a single scalar field theory with non-minimal coupling $\xi \phi^2 R$ and a quadratic term $ R^2$. 
\end{itemize} 
Our findings allow us to verify whether
\begin{enumerate}
    \item the equations of motion have a solution such that all fields approach the false vacuum at  infinity;
    \item this solution has well-defined and finite on-shell action;
\end{enumerate}
and, thus, if the vacuum decay process is hindered in the aforementioned modified gravity theories. While these conditions constrain the Coleman-de Luccia bounce, they do not apply to static solutions such as the Hawking-Moss instanton, which contributes to vacuum decay when the false vacuum has a de Sitter geometry \cite{Hawking:1981fz}. The violation of Condition 1. means that only bubbles of infinite radius are critical, that is the only solution satisfying the boundary conditions is the false vacuum static solution and thus there is no phase transition to the true vacuum. If, instead, the on-shell action is infinite, the decay rate vanishes (the so-called ``vacuum quenching'' \cite{Coleman:1980aw}). The bounce action may also be ill-defined near the false vacuum, i.e. at the upper bound of integration: this happens, for example, if our candidate metastable state is a minimum of the Euclidean potential, and thus a maximum in Minkowski space\footnote{Despite this result seems trivial, as there is no potential barrier through which the scalar field can tunnel, early studies of the vacuum decay phenomenon actually focused on tunneling without barriers \cite{lee}. Scalar field decay in a quartic potential with negative coupling is an example of this behaviour.}. Such a state does not exhibit any metastability, but only an instability related to the local unboundedness of the potential around such a fixed point. As we will see, quadratic gravity falls precisely in this class of theories.
The structure of the paper is the following.  In Sec.\ \ref{sec:scalarfield} we introduce our method by first analyzing a single scale field theory without gravity and then by including standard General Relativity. In Sec.\ \ref{sec:scaledep} we include quadratic corrections and a non-minimal coupling. In Sec.\ \ref{sec:scaleinv} we repeat our analysis without the Einstein-Hilbert term, i.e. with a scale-invariant gravitational sector. The asymptotic bounce has also numerical implications that are explored in Sec.\ \ref{sec:num}. We finally conclude in Sec.\ \ref{conclusions} we some remarks and future directions. To keep the discussion as simple as possible, we confine the most technical details in the appendices.

\section{Scalar asymptotic bounce with Einstein-Hilbert Gravity}
 \label{sec:scalarfield}
 \subsection{Scalar asymptotic bounce}
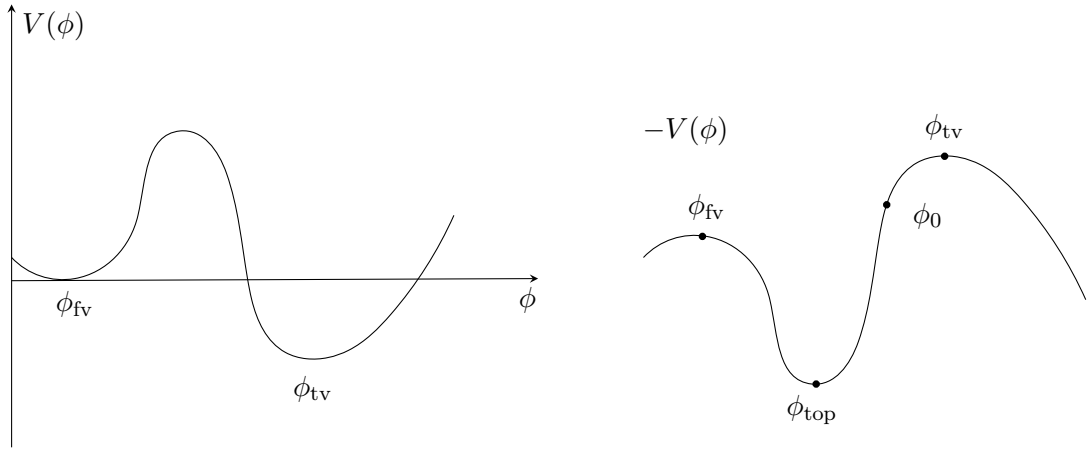
\begin{figure*}
\centering
\mbox{
\begin{tikzpicture} [scale=2.8]
\draw (0,0.1) to [curve through={(0.3,0)..( 0.6,0.3)..(0.8,0.7)..(1.02,0.5)..(1.12,0)..(1.14,-0.1)..(1.3,-0.35)..(1.4,-0.38)..(1.6,-0.34)..(1.8,-0.17)}](2.1,0.3);
\draw[-stealth] (0,-0.8) to (0,1.3);
\draw[-stealth] (0,-0.01) to (2.5,0);
\draw (3,0.1) to [curve through={(3.3,0.2)..( 3.6,-0.1)..(3.8,-0.5)..(4.02,-0.3)..(4.12,0.2)..(4.14,0.3)..(4.3,0.55)..(4.4,0.58)..(4.6,0.54)..(4.8,0.37)}](5.1,-0.1);
\filldraw[color=black, fill=black,very thick](3.28,0.2) circle (.01);
\filldraw[color=black, fill=black,very thick](3.82,-0.5) circle (.01);
\filldraw[color=black, fill=black,very thick](4.155,0.35) circle (.01);
\filldraw[color=black, fill=black,very thick](4.43,0.58) circle (.01);
\node at (0.2,1.2) [scale=1] {$V(\phi)$};
\node at (3.2,.7) [scale=1] {$-V(\phi)$};
\node at (2.45,-0.1) [scale=1] {$\phi$};
\node at (0.3,-0.12) [scale=1] {$\phi_{\rm fv}$};
\node at (1.43,-0.52) [scale=1] {$\phi_{\rm tv}$};
\node at (3.3,0.34) [scale=1] {$\phi_{\rm fv}$};
\node at (4.43,0.72) [scale=1] {$\phi_{\rm tv}$};
\node at (4.35,0.3) [scale=1] {$\phi_{0}$};
\node at (3.8,-0.62) [scale=1] {$\phi_{\rm top}$};
\end{tikzpicture}
}
 \caption{Scalar field potential with two classical vacua (left panel) and its Euclidean counterpart (right panel).  $\phi_{\rm fv}$ is the value of the scalar field at the false vacuum, separated from the local minimum of the potential by a barrier, on top of which $\phi=\phi_{\rm top}$. $\phi_0$ marks the bounce initial condition: the field undershoots (overshoots) when released for $\phi_{\rm top}\leq \phi\leq \phi_0$ ( $\phi_{0}\leq \phi\leq \phi_{\rm tv}$ ).} 
    \label{fig:potential}
\end{figure*}

 A metastable scalar field decays due to quantum tunnelling at an exponentially small rate $\Gamma$, given by  \cite{Paranjape,Andreassen:2016cvx,Coleman:1977py,Callan:1977pt}
\bea
\Gamma=A e^{-B}\,,
\eea
where $B$ is the difference between the Euclidean action calculated on the bounce and the one computed in the false vacuum state. Consider a single scalar field theory with a metastable state ($\phi_{\rm fv}$ in Fig.\ref{fig:potential}) and $O(4)-$symmetric Euclidean action
\bea
\label{eq:action}S=2 \pi^2\int_{0}^{+\infty} dt\,t^3\left(\frac{\dot\phi^2}{2}+V(\phi)\right)\,,
\eea
where the dot indicates a derivative with respect to the Euclidean radius $t$, and $\phi=\phi(t)$. For the time being, we assume also that $\phi_{\rm fv}=0$ and $V(\phi_{\rm fv})=0$ . The bounce is a solution to the equation of motion
 \bea
 \label{eq:eom}
 \ddot{\phi}+\dfrac{3 \dot{\phi}}{t}=\dfrac{dV}{d\phi}
 \eea 
 with boundary conditions
 \bea
 \label{eq:bc}
\lim_{t\rightarrow+\infty} \phi(t)=\phi_{\rm fv}\qquad \dot\phi(0)=0.
\eea
According to Eq.\ \eqref{eq:eom} the scalar field is subjected to friction and evolves as a particle moving in the potential  $-V(\phi)$. The bounce thus may be determined as the critical trajectory which separates undershoot trajectories (the scalar field does not quite reach the false vacuum, inverting its velocity somewhere between $\phi_{fv}$ and $\phi_{top}$) from overshoot ones (the scalar field reaches the false vacuum with finite velocity). 

We now derive the asymptotic behaviour of the bounce for the theory in Eq.\ \eqref{eq:action}. Let us consider a generic undershoot trajectory and look for an approximate solution for large $t\leq t^*$, where $t^*$ marks the smaller radius at which the scalar field velocity changes sign ($\dot\phi(t^*)=0$). In order to find the asymptotic bounce we then take the limit $t^*\rightarrow \infty$. The right-hand side of Eq.\ \eqref{eq:eom}, expanded around $t^*$, reads (the subscript $_*$ indicates quantities evaluated at $t^*$)
  \begin{equation}
  \dfrac{dV}{d\phi}=\left(\dfrac{dV}{d\phi}\right)_*+\sum_{n\geq2} f_{n*}\dfrac{(t-t^*)^n}{n!}\label{eq:higherorder.}\end{equation}
where the general form of the coefficients  $f_{n*}$ is reported in
   Appendix \ref{sec:appendix}. We require that every $f_n$ is such that the approximation 
  \begin{equation}  
  \label{eq:zerorder}\dfrac{dV}{d\phi}\approx \left(\dfrac{dV}{d\phi}\right)_*
  \end{equation}
  holds when $t^*$ is large \footnote{We can formulate this condition also as $t-t^*\sim -At^*$ with $A$ of order unity.}. As explained in Appendix \ref{sec:appendix}, this means that we should require
   \begin{equation}  
  \label{eq:cond1}
          \left(\dfrac{d^jV}{d\phi^j}\right)_*\ddot{\phi}_*^{j-2} t^{*2j-2}\ll1
      \end{equation} 
      for $j\geq 2$.
Under these conditions,  Eq. \eqref{eq:eom} becomes
 \begin{equation}\ddot{\phi}+\dfrac{3\dot{\phi}}{t}=\left(\dfrac{dV}{d\phi}\right)_*
 \end{equation}
and the solution reads
 \begin{equation}\phi(t)=\phi_*-\left(\dfrac{dV}{d\phi}\right)_*\dfrac{t^{*2}}{4}+ \left(\dfrac{dV}{d\phi}\right)_*\dfrac{t^2}{8}+\left(\dfrac{dV}{d\phi}\right)_*\dfrac{t^{*4}}{8 t^2}\,,
 \end{equation} 
where we used $\phi(t_{*})=\phi_{*}$ and $\dot\phi(t_*)=0$ to fix the two integration constants. Taking the limit $\phi_*\rightarrow 0$ and $t^*\rightarrow +\infty$ we find the asymptotic behaviour of the bounce
 \bea\non
\lim_{\underset{t^*\rightarrow+\infty}{\phi_*\rightarrow0}}\dot{\phi}(t)&=&-\dfrac{C_0}{t^3}\,,\quad\lim_{\underset{t^*\rightarrow+\infty}{\phi_*\rightarrow0}}\left(\dfrac{dV}{d\phi}\right)_*=\dfrac{4 C_0}{t^{*4}}\,,\\
 \lim_{\underset{t^*\rightarrow+\infty}{\phi_*\rightarrow0}}\phi(t)&=&\dfrac{C_0}{2 t^2}\,,\label{eq:behaviourm}
 \eea
 from  sufficiently large $t$ up to $t\rightarrow+\infty$. The constant $C_{0}$ is determined by the second limit  in Eqs. \eqref{eq:behaviourm}. As \bea
      \lim_{t^*\rightarrow+\infty} \ddot{\phi}_* t^{*4}=4 C_0
      \eea we should also impose the conditions
      \begin{equation}
 \label{eq:massless}
    \left( \dfrac{d^2V}{d\phi^2}\right)_* t^{*2}\ll1\,,\qquad 4 C_0\left(\dfrac{d^3V}{d\phi^3}\right)_*\ll1\,,
    \end{equation}
    to guarantee that the constraint Eq.\ \eqref{eq:cond1} holds, implying that the scalar field should be massless with small cubic self-interactions. By using Eq.\ \eqref{eq:behaviourm} it is trivial to verify that Condition 2. holds (while Condition 1. does not apply since we have only one equation of motion). The asymptotic $t^{-2}$ behaviour of the scalar field near the bounce in Eq. \eqref{eq:behaviourm} has been already observed by \cite{Bentivegna:2017qry} in the case of Higgs decay with Einstein-Hilbert gravity.  
    
      A massive but light scalar field satisfies Eq.\ \eqref{eq:behaviourm} for  $t^*\ll m^{-1}$ while for $t^*\gg m^{-1}$ instead it should be proportional to $e^{-mt}$ \cite{Affleck:1980mp}. The proportionality constant may be determined with our method, but a finite mass makes calculations much more involved, as Eq.\ \eqref{eq:massless} suggests that the potential is important at each order in the Taylor expansion. 
    
\subsection{Asymptotic bounce with Einstein-Hilbert gravity}
 The above results  can be readily generalized in the case of a single scalar field theory with Einstein-Hilbert gravity, when the false vacuum geometry is flat. The Euclidean action is 
 \bea S=\int d^4x \sqrt{g} \left[-\dfrac{M_{\rm P}^2 R}{2}+\frac12(\partial\phi)^{2}+V(\phi)\right]\,.
  \eea
 If the line element is $O(4)$-symmetric (in analogy with the flat space case Eq.\eqref{eq:action})
 \bea
 \label{eq:lineEH}
 ds^2=dt^2+\rho(t)^2 d\Omega_3^2\,,
 \eea 
 Eq. \eqref{eq:actionEH} becomes
 \bea\label{eq:actionEH}S=2\pi^2\int_0^{+\infty} dt \rho(t)^4 \left[-\dfrac{M_{\rm P}^2 R}{2}+\frac{\dot\phi^{2}}{2}+V(\phi)\right].\,
  \eea
 The equations of motion are
 \bea \label{eq:eom1EH}
&&\ddot{\phi}+3\,\dfrac{\dot{\rho}\,\dot{\phi}}{\rho}=\dfrac{dV}{d\phi}\,,\\\non\\
 &&\dot{\rho}^2=1+ \dfrac{\rho^2}{3M_{\rm P}^2} \left(\dfrac{\dot{\phi}^2}{2}-V(\phi)\right)\,.\label{eq:eom2EH}
 \eea
 As $V(\phi_{\rm fv})=0$, we have that  Eq.\ \eqref{eq:eom2EH} yields
 \bea\label{eq:approxrho}\rho(t)\approx t+\text{higher orders}
 \eea
near the bounce at large times, thus readily giving Eq.\ \eqref{eq:behaviourm} for a single scalar field theory with Einstein-Hilbert gravity. Note that  Eq.\ \eqref{eq:behaviourm} along with Eq.\ \eqref{eq:approxrho}, together with Eq. \eqref{eq:eom2EH}, proves that Condition 1. holds. Moreover the action \eqref{eq:actionEH}, computed on the asymptotic bounce, is a convergent integral at $t\rightarrow+\infty$.

\begin{figure*}
\centering
      \mbox{
     \begin{minipage}{0.5\textwidth}
     \qquad \includegraphics[scale=0.5]{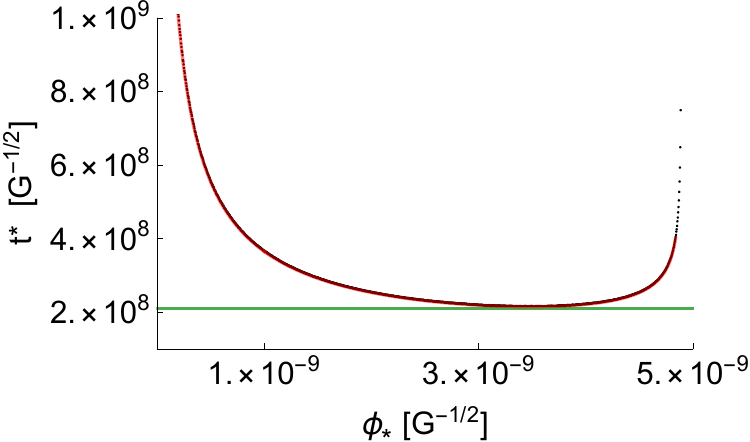}
     
     \includegraphics[scale=0.8]{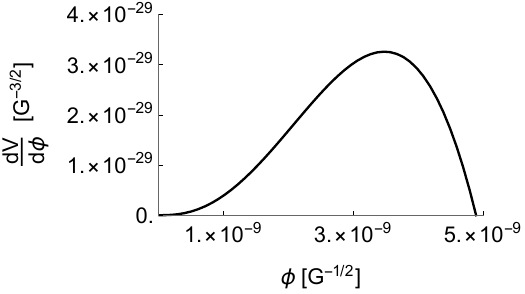}
     \end{minipage}%
     \begin{minipage}{0.5 \textwidth}
  
\quad  \includegraphics[scale=0.5]{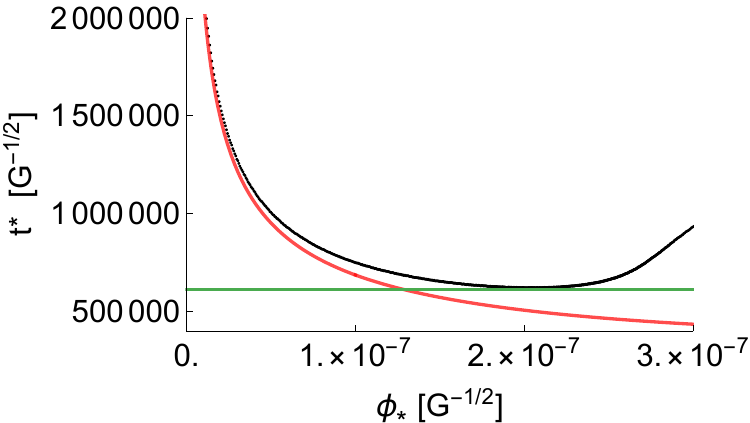}
     
     \includegraphics[scale=0.73]{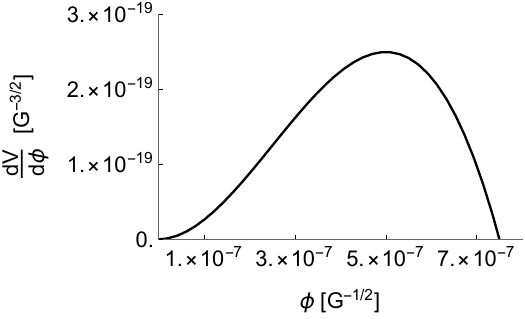}
     \end{minipage}
     }
     \caption{Top: $t^*$ as a function of $\phi_*$ for potentials Eq.\eqref{eq:higgs} (on the left) and Eq.\eqref{eq:poly} (on the right), in the vicinity of the bounce ($\phi_*\rightarrow 0$). Bottom: $V'(\phi)$ as a function of $\phi$.}
     \label{fig:mintm}
 \end{figure*}

   \subsection{Examples}\label{Examples}
 We now test our findings in two examples \footnote{the mass unit is $G= (M_{\rm P}^2 8 \pi)^{-1}=1$}:
\begin{itemize}
     \item the Higgs potential
\begin{equation}
  \label{eq:higgs}
  V(\phi)=\dfrac{\lambda(\phi)}{4} \phi^4
  \end{equation}
  where \begin{equation}\ell(\phi)=\lambda^*+\alpha' \ln\left(\phi\right)^2+\beta \ln (\phi)^4\end{equation}
 and $\lambda^*=-0.0013, \alpha'=1.4\times10^{-5},\beta=6.3 \times10^{-8}$ (see \cite{Burda:2016mou});
    \item a polynomial potential with vanishing quadratic term
    \begin{equation}
    \label{eq:poly}
        V(\phi)=\alpha_1 \phi^5+\alpha_2\phi^4+\alpha_3 \phi^3\end{equation}
        where we choose $\alpha_1=1$, $\alpha_2=-1$, $\alpha_3=10^{-6}$.
\end{itemize} 
One can easily prove that  Eq.\ \eqref{eq:cond1} is satisfied for $\phi_*\rightarrow 0$  in the Higgs case for all $\alpha',\beta,\lambda^{*} $ and for $\alpha_3 C_0\ll 1$ for the polynomial potential.
In both cases, we compare $t^*$ as a function of $\phi_*$ as given by our theoretical prediction  Eq.\ \eqref{eq:behaviourm}  (red line in Fig.\ref{fig:mintm}) with a numerical evaluation (black dots) and found good agreement among the two. The green line instead marks the position of the minimum for the numerical calculation, which corresponds to the maximum of $\dfrac{dV}{d\phi}$ in the Higgs case. We took $C_0$s to be the ones determined numerically with the method described in Sec.\ \ref{sec:num} and reported in Table\ \ref{tab:1}.

\section{The asymptotic bounce in modified gravity}
 \label{sec:scaledep} As mentioned in the Introduction, the decay of a metastable state in single scalar field theories with modified gravity may be hindered by the absence of a bounce with finite on-shell action. In this section, we consider Einstein-Hilbert gravity, a non-minimal coupling $\xi \phi^2 R$, and a quadratic Ricci scalar $\alpha R^2$ and test for Conditions 1. and 2. in each case.  These terms are usually required by renormalizability in quantized field theories on a gravitational (classical) background \cite{birrell,parker,Buchbinder:1992rb}. The Euclidean action is (the line element is again given by Eq.\ \eqref{eq:lineEH})
 \bea\non
 \label{eq:actionscale}
     S=2\pi^{2}\int_0^{+\infty}dt\rho(t)^3\Bigg(\dfrac{\dot{\phi}^2}{2}+ V(\phi)-\dfrac{M_{\rm P}^2}{2} R -\dfrac{\xi}{2} \phi^2 R+\dfrac{\alpha}{36} R^2\Bigg)
 \eea 
 where
 \bea
 \label{eq:Rscale}
 R=\dfrac{-6(\dot{\rho}^2-1+\rho\ddot{\rho})}{\rho^2}.\eea
  The equations of motion are
 \begin{gather}\label{eq:eom1scale}\dot{\rho}^2=1+\rho^2 \, \dfrac{\dfrac{\dot{\phi}^2}{2}-V(\phi)+\dfrac{\alpha}{36} R^2+\left(\dfrac{\alpha}{3}\dot{R}-6 \xi\,\phi\,\dot{\phi}\right)\dfrac{\dot{\rho}}{\rho}}{3\left(M_{\rm P}^2+\xi \phi^2-\dfrac{\alpha}{9} R\right)},\\[5pt]
 \label{eq:eom2scale}\ddot{\phi}+3\dfrac{\dot{\rho}\,\dot{\phi}}{\rho}=\dfrac{dV}{d\phi}-\xi \phi R.\end{gather}
 In addition, we consider the trace of the Einstein equation
 \bea
    0=-\left[3 M_{\rm P}^2+ 3 \xi (1+6 \xi)\phi^2+\alpha\Box\right] R +3 \dot{\phi}^2 (1+6 \xi)+12 V(\phi)+18 \xi \phi \dfrac{dV}{d\phi}. \label{eq:tracescale}
 \eea
We first consider the non-minimal coupling case ($\xi\neq0$, $M_P\neq 0$ and $\alpha=0$) , which has been already extensively analyzed, in particular concerning Higgs decay \cite{Rajantie:2016hkj,Branchina:2019tyy,Czerwinska:2016fky,Salvio:2016mvj}. No obstruction according to Conditions 1. and 2.  has been found so far, thus the same should hold for generic massless fields, since, as we have demonstrated, the asymptotic bounce is fairly independent of the potential. Still, we should verify whether Eq.\ \eqref{eq:cond1} is sufficient for Eq.\ \eqref{eq:behaviourm} to hold. Moreover, the asymptotic bounce itself may differ from the $\xi=0$ case.  We will turn then to quadratic gravity.
 
\subsection{Non-minimal coupling only}
As $\alpha=0$, $\phi$ is the only propagating scalar degree of freedom, and Eq.\ \eqref{eq:eom2scale} depends on gravity both through the friction term and the Ricci scalar.  The former is approximately given by Eq.\ \eqref{eq:approxrho} near the bounce at large times. Using Eq.\ \eqref{eq:tracescale} in Eq.\ \eqref{eq:eom2scale} and expanding around the false vacuum ($\phi_{\rm fv}=0$) we get
\bea
\label{eq:expW}
\ddot{\phi}+3\dfrac{\dot{\phi}}{t}=\dfrac{dV}{d\phi}- \dfrac{\xi}{M_{\rm P}^2} \phi\dot{\phi}^2 (1+6 \xi).
\eea
Now we expand  Eq.\ \eqref{eq:expW} around the turning point $t^*$, evaluate it for $t\leq t^*$, take the large $t^*$ limit and see under which conditions the zeroth-order term dominates the Taylor expansion, and, thus, when Eq.\ \eqref{eq:behaviourm} holds. Barring numerical cancellations\footnote{Here we are using the fact that the second derivative of $\phi^3$ contains $\phi\dot{\phi}^2$. We can trade one for the other because our calculations were carried out without accounting for numerical factors, and thus they are independent of possible numerical cancellation among different terms generated by $\phi^3$ in the Taylor expansion.}, we can write
  \bea
  \label{eq:derexp}
\sum_{n=0}^{+\infty}\left(\phi\dot{\phi}^2\right)_*\dfrac{(t-t^*)^n}{n!}\approx\sum_{n=2}^{+\infty}\left(\phi^3\right)_*\dfrac{(t-t^*)^{n-2}}{n!}
\eea
apart from numerical factors.  This gives
 \bea
\ddot{\phi}+3\dfrac{\dot{\phi}}{t}&\approx&\sum_{n=0}^{+\infty}\left(\dfrac{dV}{d\phi}\right)_*\dfrac{(t-t^*)^n}{n!}-\dfrac{\xi(1+6\xi)}{M_{\rm P}^2}\sum_{n=2}^{+\infty}\left(\dfrac{\phi^3}{t^{*2}}\right)_*\dfrac{(t-t^*)^{n}}{n!}
\eea
if we take $t-t^*\approx - A t^*$, where $A$ is a constant of order unity.  As quartic interactions always satisfy Eq.\ \eqref{eq:cond1}, we conclude that such term does not give an appreciable contribution to the Taylor expansion at large times on the bounce. Thus, if Eq.\ \eqref{eq:massless} hold on $V(\phi)$, we can safely approximate the potential as Eq.\ \eqref{eq:zerorder}, and thus Eq.\ \eqref{eq:behaviourm} holds.
Notice that taking Eq.\ \eqref{eq:behaviourm} and $\rho(t)\approx t$ in Eq.\ \eqref{eq:eom1scale} gives consistently $\dot{\rho}\approx 1$, and the Lagrangian decays sufficiently fast so that its integral Eq.\ \eqref{eq:actionscale} converges for $t\rightarrow\infty$. Thus, as expected, no obstructions to decay are found in this case.
The same calculations may be carried out in the $\phi_{\rm fv}\neq 0$ case  and lead to analogous results, with a shifted integration constant
\bea
\lim_{\underset{t^*\rightarrow+\infty}{\phi_*\rightarrow0}}\left(1-\dfrac{6 \xi^2 \phi_{\rm fv}^2}{M_{\rm P}^2+\xi (1+6\xi \phi_{\rm fv}^2)}\right)\left(\dfrac{dV}{d\phi}\right)_*=\dfrac{4 C_0}{t^{*4}}.\eea

 \subsection{Quadratic gravity only}
We now set $\xi=0$, $\alpha\neq 0$. The scalar field equation of motion depends on gravity only through the friction term, while the trace equation Eq.\ \eqref{eq:tracescale} gives the dynamics of $R$ when subjected to a scalar field source, given by the trace of the stress-energy tensor.  In this case, we have two propagating degrees of freedom, the scalar field and the Ricci scalar. As we have coupled equations of motion, it seems that we cannot use the same argument as above, since there is no clear undershoot/overshoot distinction in this case. However, one finds that the scalar field equation of motion actually decouples near the bounce at all times. In fact, if Eq.\ \eqref{eq:tracescale} holds at some time $t$, we can determine $R$ as given by Eq.\ \eqref{eq:tracescale} with
  \bea
  \label{eq:box}
\Box\approx\dfrac{d^2}{dt^2}+\dfrac{3}{t}\dfrac{d}{dt} .
  \eea
  The Ricci scalar depends only on deviations from the flat space solution (see Eq.\ \eqref{eq:Rscale}) and so, if small at some time, it should remain as such: in order to trigger large deviations from flat space in $R$, they should first appear in Eq.\ \eqref{eq:box}, but this means that higher-order deviations from flat space (in $R$), are determined by lower-order deviations (in the friction term), which is impossible.  Thus, we can safely take Eq.\ \eqref{eq:approxrho} to hold in Eq.\ \eqref{eq:eom1scale} at all times. To determine the Ricci scalar asymptotic bounce, we solve  Eq.\ \eqref{eq:tracescale} with Eq.\ \eqref{eq:box}.  We find $R=R_{hom }+R_{ps}$ with 
  \bea \label{eq:quadsol2}
     R_{\rm hom}&=&\epsilon^2C_1\,\dfrac{J_1\left(\epsilon\, t'\right)}{t}\,, \label{eq:quadsol3}\\\non
    R_{\rm ps}&=&-\epsilon^2\dfrac{ \pi}{2}\dfrac{J_1\left(\epsilon\,t'\right)}{t}\int^t F(y) Y_1\left(\epsilon^3\,y'\right) y^2 dy\\\non
     &+&\epsilon^2\dfrac{\pi}{2}\dfrac{Y_1\left(\epsilon^3\,t'\right)}{t}\int^t F(y) J_1\left(\epsilon\,y'\right) y^2 dy\,.
     \eea
     Here, $J,Y$ are Bessel function of the first kind,
 \bea
 A=\sqrt{\dfrac{3M_{\rm P}^2}{|\alpha|} }\,,\quad t'=A (t+a)\,,\quad \begin{cases} \epsilon=1&\alpha>0\\\epsilon=i&\alpha<0\,, \end{cases}
 \eea
 $C_{1,2}$ are constants, and the function $F(t)$ is given by
 \bea
 F(t)=\dfrac{3\dot{\phi}(t)^2}{\alpha}+\dfrac{12}{\alpha} V(\phi(t))\,.
 \eea
 Using Eq.\ \eqref{eq:behaviourm} we find that the Ricci scalar can be approximated as
\bea
\label{eq:Rlargetnonm}
R=\epsilon^2(C_1+\tilde{C}_1)\,\dfrac{J_1\left(\epsilon\, At\right)}{t}+\tilde{C}_2\dfrac{Y_1\left(\epsilon^3\, At\right)}{t}+O(t^{-6})
\eea
where
\bea
\tilde{C}_1=-\dfrac{ \pi}{2}\epsilon^2\int_0^{+\infty} F(y) Y_1\left(\epsilon^3\,y'\right) y^2 dy\\
\tilde{C}_2=\dfrac{ \pi}{2}\epsilon^2\int_0^{+\infty} F(y) J_1\left(\epsilon\,y'\right) y^2 dy.
\eea
Higher order terms $O(t^{-6})$ are computed by using  the asymptotic forms of the Bessel functions $J$, $Y$ for large arguments \cite{abraham} 
\begin{gather}
\label{eq:asymptbessel}
J_{1}^{\infty}(z)=\sqrt{\dfrac{2}{\pi z}} \cos\left(z-{3\pi\over 4}\right)\\
Y_{1}^{\infty}(z)=\sqrt{\dfrac{2}{\pi z}} \sin\left(z-{3\pi\over 4}\right).\notag
\end{gather}
 The Ricci scalar is dominated by the first two terms in Eq.\ \eqref{eq:Rlargetnonm} unless  both $C_1+\tilde{C}_1$ and $\tilde{C}_2$ vanish. However, one can prove that $\tilde{C}_2$ is always non-vanishing. In fact, we have
\bea
\int^{+\infty}_0 F(t) J_1(t') t^2 dt&<&\int^{+\infty}_0 t' t^2 F(t)dt
\eea
and the right-hand side is negative definite if
\bea
\label{eq:inequality}
\int^{+\infty}_0 t^3 \left(\dfrac{\dot{\phi}^2}{2}+V(\phi)\right)dt<-\int^{+\infty}_0 t^3V(\phi)dt.
\eea
To prove that, we consider an off-shell scalar field profile which corresponds to the bounce of the same theory without gravitational interaction. Then, the left-hand side of Eq.\eqref{eq:inequality} is the bounce action of said theory, in the approximation of small gravitational backreaction. As computed in \cite{Isidori:2007vm}, this profile may be used to determine the on-shell action of the theory with also an Einstein-Hilbert term (i.e., the right-hand side)\footnote{Actually, the approach undertaken in \cite{Isidori:2007vm} is slightly different, and it considers the flat-space bounce as a background for a perturbative expansion, in order to determine the action of a scalar field theory with an Einstein-Hilbert term as
\begin{equation*}
    S=S_0+\dfrac{A}{M_P^2}+O(M_P^{-4})
\end{equation*}
where $S_0$ is the on-shell action of the flat space theory and $A$ is some real constant. Some concerns have been raised in the literature \cite{Rajantie:2016hkj,Branchina:2016bws}  about this point, which is furtherly addressed in Appendix \ref{sec:appendix:pertexp}.}. One finds  \cite{Isidori:2007vm} that it is always larger than the flat space one, making $\tilde{C}_2<0$ for $\alpha>0$ and $\tilde{C}_2>0$ for $\alpha<0$. \\

Finally, using Eq. \eqref{eq:asymptbessel} in Eq.\ \eqref{eq:Rlargetnonm} shows that $R$ diverges for  $t\rightarrow+\infty$ for $\alpha<0$, thus implying that Condition 1. is violated. For $\alpha>0$, instead,  $R$  undergoes damped oscillations around the fixed point $R=0$. Such oscillations are, to the leading order, the same of a free massive scalar field around its minimum, and they make the action undefined in the upper limit of integration, thereby violating Condition 2. 
It seems thus that adding a  gravitational degree of freedom does not lead to vacuum decay from the state $\phi=\phi_{\rm fv}$, $R=0$, independently on the value of $\alpha$. This is related to the boundary value nature of the problem: it makes the bounce solutions non-perturbative as new interactions or additional scalar fields are turned on\footnote{This is analogous to what happens to massive scalar fields with quartic self-interactions. If the mass vanishes, there is a bounce given by
\begin{equation*}
    \phi(t)=\sqrt{\dfrac{2}{\lambda}}\dfrac{2\mathcal{R}}{t^2+\mathcal{R}^2}
\end{equation*}
where $\mathcal{R}$ is the bounce radius $\phi_0=2 \mathcal{R}$ but it disappears as soon as the mass is turned on.}. In the present case the stress-energy tensor naturally couples gravity and matter, making the new (gravitational) degree of freedom necessarily interacting with the scalar field. One may then ask if the obstruction arises merely by the addition of the new degree of freedom or, instead, also by non-derivative terms in Eq.\eqref{eq:tracescale}. If the former holds, one might guess that a bounce is forbidden also in theories with a scalar field, an Einstein-Hilbert term and a generic coupling $\alpha R^n$, with integer $n$, as they also have two degrees of freedom.  Eq.\eqref{eq:tracescale} in that case is
\bea
\label{eq:fullrn}
\alpha n (n-1) R^{n-3}\left(R \ddot{R} +\dfrac{3\dot\rho}{\rho}R\dot R +(n-2) \dot R^2 \right)=-6 M_p^2 R+6 \dot{\phi}^2 +24 V(\phi).
\eea
Near the bounce at large times one has
\bea
\left(\ddot R+\dfrac{3\dot \rho}{\rho}\dot R\right) R^{n-2}\ll R
\eea
for $n>2$ and $\dot R^2 R^{n-3}\ll R$ for $n>3$. Then, for $n>3$, $R$ is approximately given by 
\bea
R\approx\dfrac{\dot{\phi}^2 +4 V(\phi)}{M_P^2}
\eea
 which does not allow to exclude in principle a bounce. Then, obstructions related to Conditions 1. and 2. that arise in the $n=2$ case seems not to be merely related to the presence of derivatives of $R$ in the Einstein trace equation. For $n=3$ instead one has
\bea
\label{eq:approxrn}
\alpha\dot R^2=-M_p^2 R+\dot{\phi}^2 +4 V(\phi)
\eea
near the bounce at large times. Now one would need to determine which terms dominate Eq.\eqref{eq:approxrn} and thus the asymptotic bounce of $R$. Our calculations in the Appendix rely though on a specific form of the equation of motion. Moreover, we weren't able to find a full solution to Eq.\eqref{eq:fullrn}. This makes a solution not feasible with our present means.

\subsection{Non-minimal coupling and quadratic gravity}
We have seen that including a squared Ricci scalar results in a bounce action that is ill-defined for $\alpha>0$. Thus, we expect that setting $\xi\neq 0$, along with $\alpha\neq 0$ and $M_{\rm P}^2 \neq 0$, does not change much the situation and we now show that this is the case.   Eq.\ \eqref{eq:eom2scale} and Eq.\ \eqref{eq:tracescale} are coupled, so we need first to disentangle them to read the scalar field asymptotic bounce and use it to find the Ricci scalar. To do that, we use Eq.\ \eqref{eq:eom2scale} in Eq.\ \eqref{eq:tracescale} to replace non-derivative terms in  $R$. We find
 \bea
 \label{eq:solnonm}
 R=C_1+\int^t t'^{-3} \int^{t'} F(t'') t''^{3} dt'dt''
 \eea
 with
 \bea
\alpha  F(t)&=&3 \dot{\phi}^2 (1+6 \xi)+12 V(\phi)+18 \xi \phi \dfrac{dV}{d\phi}\\\non
&-&3 \xi \left( (1+6 \xi) \phi+\dfrac{M_{\rm P}^2}{\phi}\right)\left(\ddot{\phi}+3\dfrac{\dot{\phi}}{t}-\dfrac{dV}{d\phi}\right).
      \eea
Defining $f(t)$ as
\bea
\label{eq:defft}
\int^t t'^{-3} \int^{t'} F(t'') t''^{3} dt'dt''\equiv f(t) F(t)
\eea
and replacing $R$ in Eq.\ \eqref{eq:eom2scale} with Eq.\ \eqref{eq:solnonm} we find
 \begin{widetext}
 \bea
 \label{eq:eommix}
 \ddot{\phi} F_1(\phi,t)+\dfrac{3\dot{\phi}}{t} F_2(t)=\dfrac{dV}{d\phi}F_3(t)+12 \xi \phi V(\phi) f(t)
 \eea
 with
 \bea
 F_1(\phi,t)=1+\dfrac{3\xi}{\alpha
} \phi f(t) \left( (1+6 \xi) \phi+\dfrac{M_{\rm P}^2}{\phi}\right)\approx \dfrac{3\xi}{\alpha
}  M_{\rm P}^2 f(t) +1
 \eea
  \bea
 F_2(\phi,t)=1-\dfrac{3\xi}{\alpha
}  (1+6\xi) \phi \dot{\phi}t f(t)+\dfrac{3\xi}{\alpha
} \phi f(t) \left( (1+6 \xi) \phi+\dfrac{M_{\rm P}^2}{\phi}\right)\approx \dfrac{3\xi}{\alpha
}  M_{\rm P}^2 f(t)+1
 \eea
   \bea
   F_3(\phi,t)=1+\dfrac{18\xi^2}{\alpha
} \phi^2 f(t)+\dfrac{3\xi}{\alpha
} \phi f(t)\left( (1+6 \xi) \phi+\dfrac{M_{\rm P}^2}{\phi}\right)\approx  \dfrac{3\xi}{\alpha
}  M_{\rm P}^2 f(t) +1
 \eea
 \end{widetext}
 near the bounce at large times. Therefore, the scalar field equation of motion may be approximated as 
 \bea
 \ddot{\phi}+\dfrac{3 \dot{\phi}}{t}=\dfrac{dV}{d\phi}
 \eea
 for small $\phi$, as if the non-minimal coupling was negligible. Thus we find again that $\phi(t)\propto t^{-2}$, which leads to the same obstructions to vacuum decay as in the previous case.

 \section{Scale-invariant gravity}
\label{sec:scaleinv}
 In this section, we consider a scale-invariant gravitational sector (that is we set $M_{\rm P}=0$), leaving $V(\phi)$ as the only possible source for mass scales in the theory.  From a cosmological perspective, an extremely rich phenomenology arises if also the scalar field sector is scale-invariant  
\cite{Rinaldi:2015uvu,Bezrukov:2012hx,Wetterich:2019qzx}, therefore it is important to study the stability of these configurations against vacuum decay. As we will see, our findings regarding the asymptotic bounce consistently differ from the ones found in the previous section. 

\subsection{Non-minimal coupling}
\begin{figure*}
    \centering\mbox{\begin{minipage}{0.5\textwidth}
    \includegraphics[scale=0.56]{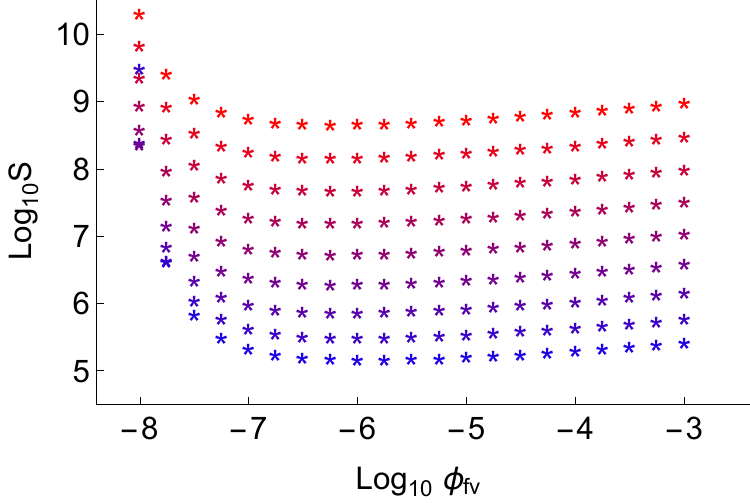}
    \end{minipage}
    \begin{minipage}{0.5\textwidth}
     \includegraphics[scale=0.6]{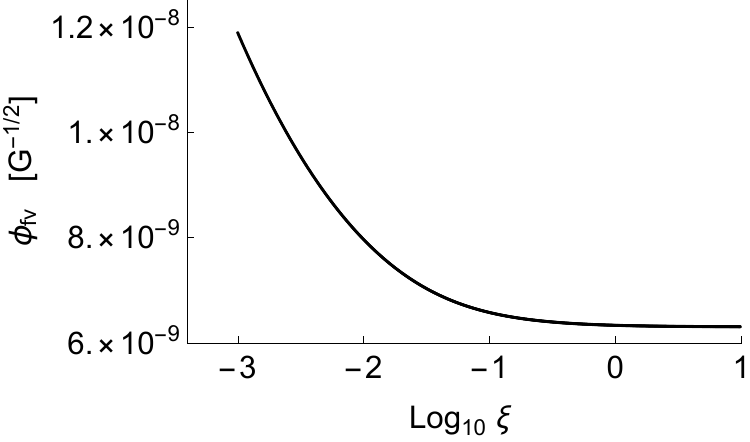}
    \end{minipage}}
    \caption{Left: numerical bounce action as a function of  $\phi_{\rm fv}$ (in units $G=1$). The non-minimal coupling is changed from $\xi=0.01$ (red) to $\xi=10$ (blue). The action sharply increases for $\phi_{\rm fv}\approx 10^{-8}$. The bounce disappears for lower values of $\phi_{\rm fv}$. Right: zeros of $F(\phi_{\rm fv},\xi)$  as a function of $\xi$.}
    \label{fig:fvcond}
    \end{figure*}

We begin with the simplest case, when also $\alpha=0$.  By combining Eq.\ \eqref{eq:tracescale} with Eq.\ \eqref{eq:eom2scale} to eliminate $R$, we find an analogous equation to Eq.\ \eqref{eq:eom}, namely
  \bea\label{eq:fieldsquared}
  \ddot{u}+3\dfrac{\dot{\rho}\,\dot{u}}{\rho}=\dfrac{dW}{du}\equiv\dfrac{4}{1+6\xi}\left(u \dfrac{dV}{du}-2 V(u)\right)\,,\eea
  where  $u\equiv\phi^2$. 
As for Eq.\ \eqref{eq:eom}, if $u$ is massless with sufficiently small cubic interactions with respect to the potential $W(u)$, we have that the asymptotic behaviour is as in Eq.\ \eqref{eq:behaviourm} and thus 
  \bea
  \phi(t)={\sqrt{C_0}\over t}\,.
  \eea Plugging this solution in Eq.\ \eqref{eq:eom1scale} we see that there is an inconsistency in  boundary conditions for gravity, as using $\rho(t)=t$ and Eq.\ \eqref{eq:behaviourm} we find at leading order in the large $t$ limit
  \bea\label{eq:wrongrhosi}\dot{\rho}^2=3+\dfrac{1}{3\xi}-\dfrac{V(\phi)t^4}{3\xi C_0}+ O\left(t^{-1}\right).\eea 
  Thus, there is no bounce if the false vacuum lives on flat space, unless also the scalar field sector is scale-invariant, with potential \footnote{Our analysis actually excludes scale-invariant potentials as there is no undershoot/overshoot distinction in that case. The asymptotic bounce (and thus possible violations of conditions 1. and 2. ) should thus be found with other methods.}
  \bea
  V\simeq {1\over C_{0}}\left(6\xi+1\right)\phi^{4}.
  \eea
This situation changes if the scalar field has a non-vanishing false vacuum value. In fact, if the potential is such that $V(\phi_{\rm fv})=0$, $\phi_{\rm fv}\neq 0$ and  $V(u)$ satisfies Eq.\ \eqref{eq:massless}, we have that, on the bounce at large times,
  \bea
  \label{eq:finitefv}
  \phi(t)\approx\sqrt{\phi_{\rm fv}^2+\dfrac{C_0}{2 t^2}}.
  \eea
   From the discussion above we expect that the asymptotic bounce is reached only in a narrow region around $\phi_{\rm fv}$ (otherwise we would get again Eq.\ \eqref{eq:wrongrhosi}), namely
 \bea
 \label{eq:ineqsi}
 \phi_{fv}^2\gg \dfrac{C_0}{2 t^2}
 \eea
 and thus we can replace Eq.\ \eqref{eq:finitefv} with  
\bea 
  \label{eq:finitefv2}
\phi(t) \approx \phi_{\rm fv}+\dfrac{C_0}{4 \phi_{\rm fv} t^2}\,.
 \eea
  Using Eq.\ \eqref{eq:eom2scale}, Eq.\ \eqref{eq:finitefv2} and Eq.\ \eqref{eq:approxrho} we get
  \bea
  \dot{\rho}(t)=1+O\left(t^{-4}\right).
 \eea 
 Eq.\ \eqref{eq:ineqsi} implies that, if $\phi_{\rm fv}$ is much smaller than the potential barrier width $\ell$\ \footnote{The potential barrier width is defined here as the range of $\phi$ such that $V(\phi)>0$.} the scalar field has not yet reached the asymptotic bounce regime when $V(\phi)>0$. This possibly makes $\dot{\rho}$ vanish, if additionally $V(\phi)\gg \dot{\phi}^2$, making $\phi$ self-accelerated\footnote{Having $\dot\rho=0$ somewhere on the bounce forbids to satisfy the requirement $\dot\rho>0$ at large $t$. In fact, this means that values for $t$ such that $\dot\rho=0$ should come in pairs but, after the first zero, one has
 \begin{equation*}
     \dfrac{d}{dt}\left(\dfrac{\dot\phi^2}{2}\right)=\dfrac{dV(\phi)}{dt}-3\dfrac{\dot\rho\dot\phi^2}{\rho}>\dfrac{dV(\phi)}{dt}
 \end{equation*}
 and thus one cannot have another one.}.
 Imposing $\dot{\rho}=0$ for $\phi\approx \phi_{\rm fv}+\ell$ and using $V(\phi)\gg \dot{\phi}^2$ gives
   \bea
   \rho_{\ell}^2=\dfrac{3 \xi (\phi_{\rm fv}+\ell)^2}{V(\phi_{\rm fv}+\ell)}.
   \eea
   This behaviour is avoided if $\rho_{\ell}>\bar{t}$, where $\bar{t}$ indicates the time by which 
 \bea
 \dfrac{\dot{\phi^2}}{2}= V(\phi)
 \eea
 on the asymptotic bounce, which in turn gives
 \bea
 C_0= -2 \phi_{\rm fv}\bar{t}^3\sqrt{2 V(B \phi_{\rm fv}})
 \eea
 where $B$ is some constant of order unity. Setting also
 \bea
 \label{eq:transition}
 u_{\rm fv}\approx \dfrac{C_0}{2 \bar{t}^2}
  \eea
  we find the condition on $\phi_{\rm fv}$, $\ell$ and $\xi$ in order for the bounce to exist
 \bea
 \label{eq:condnonmsi}
 \dfrac{6 \xi(\phi_{\rm fv}+\ell)^2 V(B \phi_{\rm fv})}{\phi_{\rm fv}^2 V(\phi_{\rm fv}+\ell)}+1<0.
 \eea
 This condition is marginally satisfied for $\phi_{\rm fv}\approx \ell$ and $\xi\approx\dfrac{1}{24}$. Decreasing $\xi$ shrinks the range of $\phi_{\rm fv}$ for which a bounce is forbidden. Moreover, Eq.\ \eqref{eq:transition} underestimates the actual $\bar{t}$ thus having the same effect. We tested this result taking as $V(\phi)$ the Higgs potential Eq.\ \eqref{eq:higgs} with $\phi\rightarrow \phi-\phi_{\rm fv}$. 
 We computed numerically the values of $\phi_{\rm fv}$ for which  
 $F(\phi_{\rm fv},\xi)=0$, with $F(\phi_{\rm fv},\xi)$ defined as
 \bea
 F(\phi_{\rm fv},\xi)\equiv\dfrac{6 \xi(\phi_{\rm fv}+\ell)^2 V(B \phi_{\rm fv})}{\phi_{\rm fv}^2 V(\phi_{\rm fv}+\ell)}+1
 \eea
 and reported the result in Fig.\ref{fig:fvcond} (on the right) as a function of $\xi$. We found that the zero of $F(\phi_{\rm fv},\xi)$ decreases for increasing $\xi$, and they lie approximately at $\phi_{\rm fv}\approx \ell \approx 5\times 10^{-9} G^{-1/2}$. We also found the bounce numerically, varying $\phi_{\rm fv}\in [10^{-8},10^{-3}]$ and $\xi\in[0.01,10]$. The on-shell  action is reported in Fig.\ref{fig:fvcond}, on the left. We found that the action sharply increases for $\phi_{\rm fv}\geq 10^{-8}$ and there is no bounce for lower $\phi_{\rm fv}$. The value of $\phi_{\rm fv}$ for which the bounce disappears increases for increasing $\xi$. They are larger than the prediction reported in Fig.\ref{fig:fvcond} as $\bar{t}$ is actually an underestimation of the matching time. \\
 
These considerations suggest that a non-minimally coupled Higgs field has no bounce, since it has a vacuum expectation value at $v=246\text{GeV}\approx 10^{-17} G^{1/2}\ll \ell$.  Actually, $v$ is generated by the interplay of a mass term and the quartic interaction: the Higgs mass affects the asymptotic bounce and in principle changes these results. Nonetheless, much of our reasoning is focused on the behaviour near the potential barrier, which is unaffected by the mass term in a large range of field values. Still, the condition in Eq.\ \eqref{eq:condnonmsi} depends on the scalar field behaviour near the false vacuum, and thus it should be reconsidered in the massive scalar field case. This will be addressed in future work.
  \subsection{Quadratic gravity}
  We consider here the effect of quadratic gravitational terms on the bounce, by setting $M_{\rm P}=0,\, \xi=0,\, \alpha\neq 0$. As in Sec.\ \ref{sec:scaledep}, the scalar field on the bounce at large times is independent of $R$, and so, if Eq.\ \eqref{eq:cond1} hold, then we have Eq.\ \eqref{eq:behaviourm}. The solution to the trace equation when the scalar field is given by Eq.\ \eqref{eq:behaviourm}, and Eq.\ \eqref{eq:approxrho} holds, is
      \bea
      \label{eq:Rsiquadratic}
      R(t)&=&C_1+\dfrac{C_2}{2 t^2}+\dfrac{3C_0^2}{8t^4 \alpha}+12\int^t\dfrac{dt'}{t'^3}\int^{t'}dt' t^{'3} V\left(\dfrac{C_0}{2 t^2}\right)
      \eea
      with $C_1$ and $C_2$ are real constant. Using  Eq.\ \eqref{eq:Rsiquadratic} , Eq.\ \eqref{eq:approxrho} and  Eq.\ \eqref{eq:behaviourm}  in Eq.\ \eqref{eq:eom1scale} we find $\dot{\rho}\neq 1$ at large times on the bounce. This result is independent of the values of $C_1$ and $C_2$. Thus, there is no bounce for scale-invariant gravity with a quadratic Ricci scalar and flat Euclidean spacetime in the false vacuum.

   \subsection{Non-minimal coupling and quadratic gravity}
We consider now a theory with a scalar field non minimally coupled to gravity and a quadratic Ricci term, namely we set $\xi\neq 0$, $\alpha\neq 0$, $M_{\rm P}=0$, and repeat the calculations of the $M_{\rm P}\neq 0$ case.
In order to have a finite bounce action $f(t)$ should satisfy
\bea
\lim_{t\rightarrow+\infty}  \phi^2 f(t)=0
\eea
as $F(t)$ is monotonically decreasing at large times on the bounce
\bea
f(t)&=&\dfrac{\int_t^{+\infty} t'^{-3} \int_{t'}^{+\infty} t''^{3} F(t'')}{F(t)}\\\non
&\leq& \int_t^{+\infty} t'^{-3} \int_{t'}^{+\infty} t''^{3} =\dfrac{t^2}{8}.
\eea
We find  
\bea
 F_1(\phi,t)\approx F_2(\phi,t)\approx F_3(\phi,t)\approx 1
 \eea
 so again 
\bea
 \ddot{\phi}+\dfrac{3 \dot{\phi}}{t}=\dfrac{dV}{d\phi}
 \eea
 for small $\phi$. Thus we expect that $\phi$ satisfies Eq.\ \eqref{eq:behaviourm} at large times on the bounce and (from Eq.\ \eqref{eq:solnonm})
\bea
\label{eq:solscinv}
R=\dfrac{C_2}{2 t^2}+\dfrac{3 (1+6\xi) C_0^2}{8 \alpha t^4}
\eea
assuming that the potential contributes only to higher orders in inverse powers of $t$. Plugging these solutions in Eq.\ \eqref{eq:eom1scale} we find that Condition 1. is violated. The situation might improve by considering a non-vanishing vacuum value for the scalar field $\phi_{\rm fv}\neq 0$. This would amount to adding a linear non-minimal coupling to gravity, $\phi R$, and having an Einstein-Hilbert term on the bounce at large times, given by $\dfrac{\xi \phi_{\rm fv}^2}{2}R$. The first one changes the right-hand side of Eq.\ \eqref{eq:eom2scale} by a multiplication constant, while we already say that Condition 2. is violated for $M_{\rm P}\neq 0$. Thus, we expect that there is no bounce also in this case.

 \section{Numerical implications of the asymptotic bounce}
  \label{sec:num}
 Having the asymptotic bounce at disposal allows improving existing numerical methods and introducing some new others. In this section, the possible implications of the asymptotic bounce for the shooting method are considered, and an alternative numerical method to find the bounce is discussed. 
 \subsection*{A cut-off for the shooting method}
In the shooting method, one finds the bounce numerically as the trajectory bracketed between undershoots and overshoots. This allows determining it with arbitrary precision, computational limits aside. In general, using this method implies a large range of integration, as the bounce initial condition $\phi(0)$ can be large ($O(0.1 M_P$)) and/or the friction term can be very effective in slowing down the scalar field. Moreover, one should compute the bounce with sufficient precision to get a good estimate of the on-shell action $S_E$. The Lagrangian must be integrated up to a cut-off, which should be carefully chosen. Knowing the asymptotic bounce allows for a different method to find the (numerical) on-shell action instead of truncation by matching the numerical bounce with the asymptotic one at some $\bar{t}$. The action may be computed as
  \bea
  \label{eq:sec1}
  S_E=S_{C,1}+S_{C,2}
  \eea
  with  
  \begin{gather}
     S_{C,1}\equiv2 \pi^2 \int_0^{\bar{t}} \sqrt{\bar{g}} \mathcal{L}(\bar{\phi},\bar{R})dt\\
     S_{C,2}\equiv2\pi^2\int_{\bar{t}}^\infty t^3 \mathcal{L}(\phi_{+\infty},R_{+\infty})dt
\end{gather}
where $\bar{R}$ and $\bar{\phi}$ are determined numerically and $\phi_{+\infty}, R_{+\infty}$ are respectively the scalar field and the Ricci scalar as given by the asymptotic bounce, which in our case depends on an integration constant $C_0$. $C$ and $\bar{t}$ may be determined by continuity as
 \begin{equation}\label{eq:match}
 \bar{t}+2\dfrac{ \phi(\bar{t})}{\dot{\phi}(\bar{t})}=0
 \end{equation}
  and $C=-\bar{t}^3\, \dot{\phi}(\bar{t})$. 
Overshoot trajectories satisfy Eq.\eqref{eq:match} at some finite $\bar{t}$ and then
\bea
\phi(t)<\phi_{+\infty}(t)\qquad \text{ for $t>\bar{t}$.}\eea
Instead undershoot trajectories near the bounce have 
\bea
\phi(t)>\phi_{+\infty}(t)
\eea
 \begin{table}[t]
 \small
     \centering
     
\resizebox{\columnwidth}{!}{%
     \begin{tabular}{|c|c|c|c|c|c|c|c|}
     \hline
     &$\phi_0$&$\dfrac{|\Delta S|}{S_{sm}}$&$S$&$C_0$&$\bar{t}$&$t^*$&$\dfrac{(S_J-S_E)}{S_J}$\\
     \hline
Higgs&$0.071$&$ 10^{-4}$&$2063.3$&$17.6\times 10^3$&$ 10^3$&$10^{10}$&$\times$\\ 

Polynomial $\alpha_3=10^{-6}$    &$24 \times 10^{-4}$ & $  10^{-4}$&$6.5961$&$1617$&$10^4$&$10^6$&$\times$\\

Polynomial $\alpha_3=10^{-5}$    & $76 \times 10^{-4}$ & $  10^{-5}$&$6.6293$&$537$&$10^3$&$10^6$&$\times$\\

Polynomial $\alpha_3=10^{-4}$    & $23 \times 10^{-3}$ & $ 10^{-3}$&$6.7442$&$172$&$10^3$&$10^5$&$\times$\\

Polynomial $\alpha_3=10^{-3}$    &  $68 \times 10^{-3}$ & $ 10^{-3}$&$7.1529$&$59.0$&$10^2$&$10^4$&$\times$\\

Polynomial $\alpha_3=10^{-2}$    &  $18 \times 10^{-2}$ & $ 10^{-3}$&$9.0185$&$23.3$&$10^2$&$10^4$&$\times$\\
Higgs + $0.1 \phi^2 R/2$&$154\times 10^{-3}$ & $10^{-5}$&$2049$&$8160$& $10^5$&$10^8$&$5\times 10^{-5}$ \\ 
  Higgs + $ \phi^2 R/2$&$17.7 \times 10^{-3}$ & $ \times10^{-5}$&$2094$&$7.04\times 10^5$& $10^5$&$10^{10}$&$3\times 10^{-3}$ \\ 
  Higgs + $10 \phi^2 R/2$&$2.00 \times 10^{-3}$ & $10^{-3}$&$2140$&$6.42 \times 10^5$& $10^6$&$10^{10}$&$3 \times 10^{-2}$ \\ 
  \hline
     \end{tabular}}
     \caption{On-shell action computed with the asymptotic bounce cut-off with the shooting method ($sm$ in the Table) one. The initial condition $\phi_0$, the on-shell action  computed with the asymptotic bounce cut-off $S$ and its relative deviation with respect to the shooting method result are reported, along with $C_0$ and the order of magnitude of $\bar{t}$ and $t^*$.}
     \label{tab:1}
 \end{table}
 \begin{figure}
 \centering
 \includegraphics[scale=0.7]{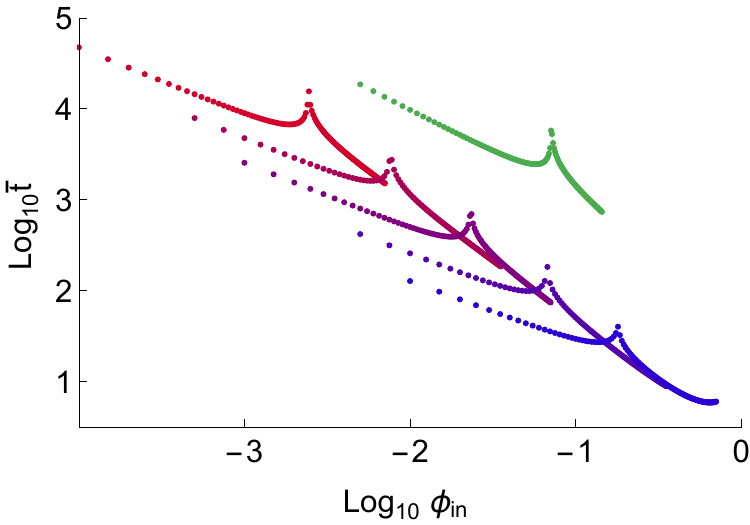}
      \caption{Matching time $\bar{t}$ as a function of the initial condition $\phi_{in}$ for a single-scalar field theory with Einstein-Hilbert gravity. The maximum marks the bounce initial condition $\phi_0$. Plots from red to blue corresponds to theories with potential Eq.\ \eqref{eq:poly} (from $\alpha_3=10^{-6}$ to $\alpha_3=10^{-2}$), the green one for the Higgs theory with Einstein-Hilbert gravity (the potential is Eq.\ \eqref{eq:higgs}).}
    \label{fig:bart}
\end{figure}
   \begin{table*}[t]
     \centering
     \begin{tabular}{|c|c|c|}
     \hline
     &$t_C$&$t_{C,sm}$\\
     \hline
Higgs&$374\text{s}$&$134\text{s}$\\ 

Polynomial $\alpha_3=10^{-6}$    &$350\text{s}$&$93 \text{s}$\\

Polynomial $\alpha_3=10^{-5}$   &$277\text{s}$&$102 \text{s}$\\

Polynomial $\alpha_3=10^{-4}$   &$276 \text{s}$&$76\text{s}$\\

Polynomial $\alpha_3=10^{-3}$   &$331\text{s}$&$113\text{s}$\\

Polynomial $\alpha_3=10^{-2}$   &$346\text{s}$&$211\text{s}^*$\\
  \hline
     \end{tabular}
         \caption{We report the computational time needed to find the bounce by minimization with the shooting method, in theories with a scalar field with potentials Eq.\ \eqref{eq:higgs} and Eq.\ \eqref{eq:poly}  and Einstein-Hilbert gravity.}
     \label{tab:2}
 \end{table*}
 at sufficiently large times, and thus there may be no $\bar{t}$ for which Eq.\eqref{eq:match} holds. Then, $\bar{t}$ may be determined as the point of closest approach
 \begin{equation}
 3 -2\dfrac{ \phi(\bar{t})\ddot{\phi}(\bar{t})}{ \dot{\phi}(\bar{t})^2}=0.
 \end{equation} 
 The matching time as defined here separates the bounce-like behaviour of undershots and overshots from the region in which they part, and thus it should get infinitely large on the bounce. Nonetheless, $\bar{t}$ is always finite off the bounce, and in particular $\bar{t}\leq t^*$.\\ The on-shell action $S_C$ with the asymptotic bounce cut-off and the one found with the shooting method $S_{sm}$ in scalar field theories with Einstein-Hilbert gravity and a non-minimal coupling are compared in Table\ \ref{tab:1}. We found that $\bar{t}\ll t^*$ and $\bar{t}$ has a maximum on the bounce as a function of $\phi_{in}$ (Fig.\ref{fig:bart}). There is a small relative deviation among $S_C$ and $S_{sm}$ and calculations in the Jordan and in the Einstein frame are in good agreement. In the Higgs case, $C_0$ roughly corresponds to the one derived from minimization of the on-shell action, with a small backreaction \cite{Isidori:2007vm}
$$C_0=2\lim_{t\rightarrow +\infty}h(t) t^2=\dfrac{4\sqrt{2}}{|\lambda|}\mathcal{R}=17.4\times 10^5\quad\text{with}\quad\mathcal{R}=350.$$

\subsection*{An alternative numerical method}

\begin{figure}
 \mbox{
     \begin{minipage}{0.5\textwidth}
     \includegraphics[scale=0.6]{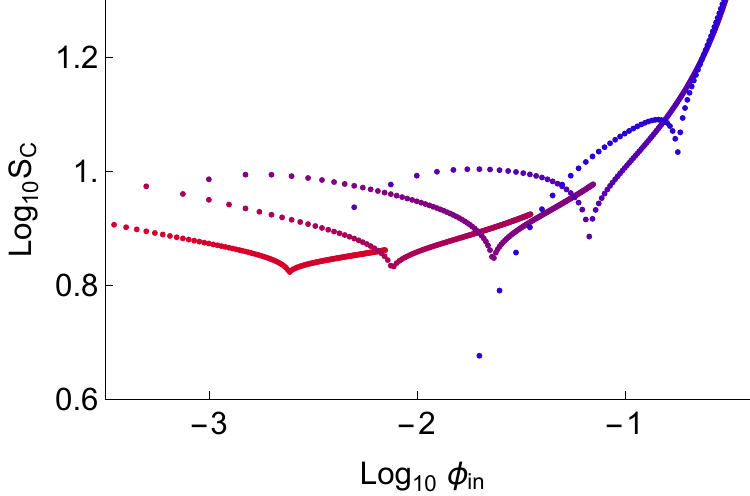}
     \end{minipage}
     
     \begin{minipage}{0.5\textwidth}
      \includegraphics[scale=0.6]{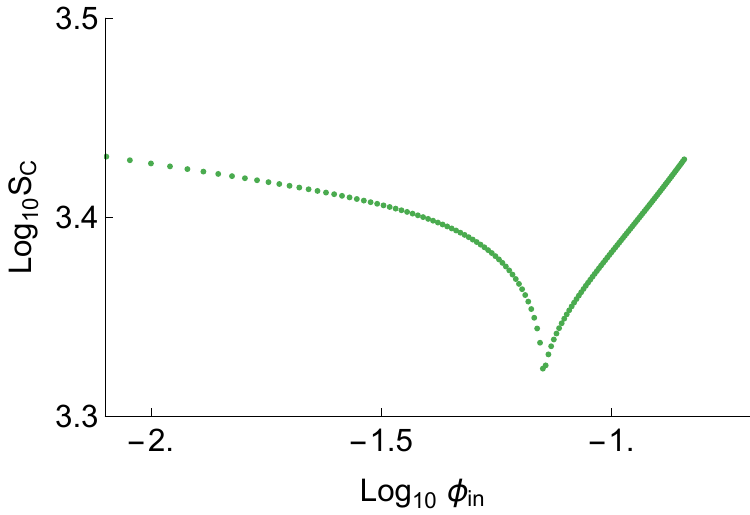}\end{minipage}}
  \caption{$S_C$ as a function of $\phi_{in}$ in units of $G$ for a single scalar field with potential Eq.\ \eqref{eq:higgs}, Einstein-Hilbert gravity and a non-minimal coupling (top left: $\xi=0.1$, top right: $\xi=1$, bottom $\xi=10$.)}
     \label{fig:sec}
 \end{figure}

\begin{figure}
     \centering
     \includegraphics[scale=0.6]{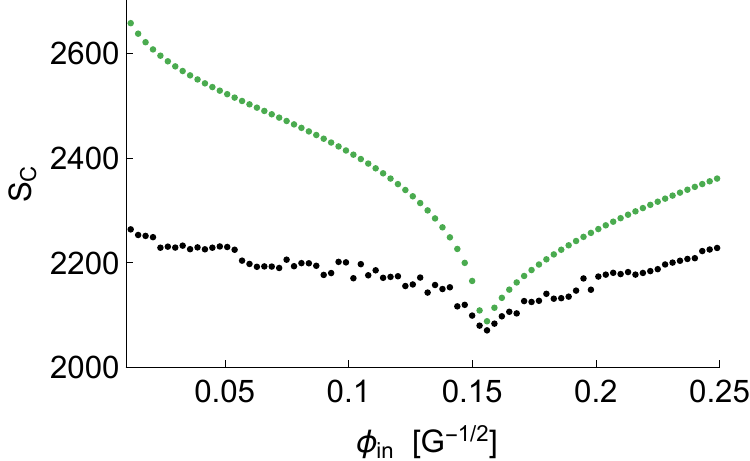}
      \includegraphics[scale=0.6]{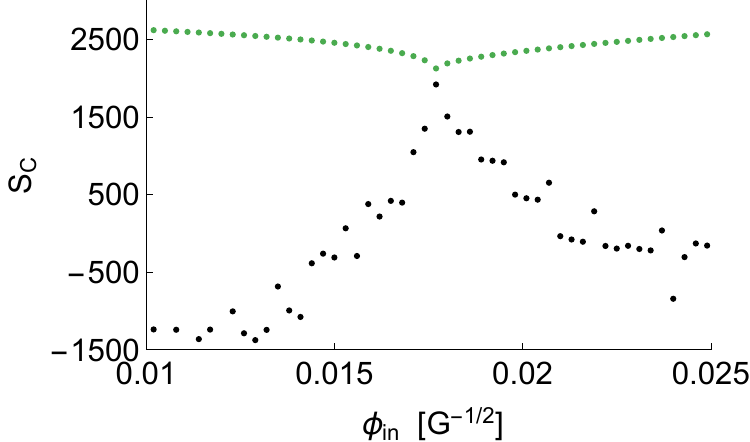}
       \includegraphics[scale=0.6]{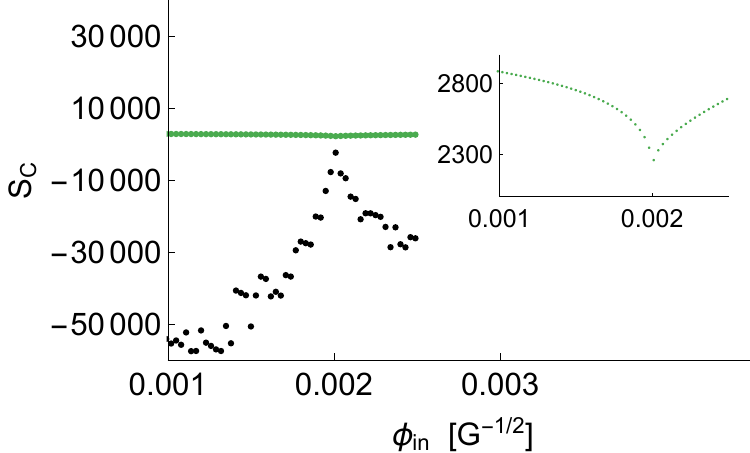}
    \caption{$S_C$ as a function of $\phi_{in}$ in units of $G$ for a single scalar field with potential Eq.\ \eqref{eq:higgs}, Einstein-Hilbert gravity and a non-minimal coupling (top left: $\xi=0.1$, top right: $\xi=1$, bottom $\xi=10$.)}
     \label{fig:secnonm}
 \end{figure}

The discussion above suggests that, for every trajectory with initial conditions
\bea
\phi(0)=\phi_{in}\qquad \dot{\phi}(0)=0
\eea
that, in general, has infinite on-shell action, there is another one that is on-shell only for $t<\bar{t}$, and that has finite $S_{C}$. $S_{C}$ should have a stationary point on the bounce, as Eq.\eqref{eq:behaviourm} holds on-shell for $\phi(0)=\phi_0$. One can show that $S_C$ has a saddle point there (see Appendix \ref{sec:appendixc}) and thus it is not suitable for minimization to find the bounce. Instead, by slightly changing this functional, one can turn the saddle point into a minimum (or a maximum), at least in the case of single-scalar field theories with Einstein-Hilbert gravity and a non-minimal coupling. The full calculation is reported in Appendix \ref{sec:appendixc}. The new functional is given by 

\bea
     &&S_{C,1}\equiv 2 \pi^2 \int_0^{\bar{t}} \bar{\rho}(t)^3 \mathcal{L}(\bar{\phi},\bar{R})dt\\\non
     &&S_{C,2}\equiv2\pi^2\int_{\bar{t}}^\infty t^3 \left(\dfrac{C^2}{2 t^6}+V\left(\dfrac{C}{2 t^2}\right)\right)dt
\eea 
 and it has a minimum on the bounce for $\xi\geq -\dfrac{1}{6}$ and a maximum otherwise (see Fig.\ref{fig:sec}-\ref{fig:secnonm}) Te computational time is of the same order of magnitude of the shooting method one (see Table\ \ref{tab:2}) which gives no clear advantage in using this method over the standard shooting method. 

 \section{Discussion}
 \label{conclusions}
 Vacuum decay is a very important test to assess the quantum stability of a field configuration with gravity but it is often very difficult to perform calculations. Most of these are numerical with poor analytical control. In this paper, we looked for an analytical method that potentially can overcome this issue and applied it to quadratic gravity with massless scalar fields.  We used it to find the large-time behaviour of the bounce (what we called the ``asymptotic bounce") provided that some restrictions on the scalar field potential (in particular a masslessness condition) hold. We showed that this result can be extended to include modified gravity scenarios too. In this way we can test for the finiteness of the action and the consistency of bounce boundary conditions for all fields, which we called Conditions 1. and 2. in the Introduction. In particular, we have studied well-founded modifications of gravity, which are required by renormalizability, namely the non-minimal coupling among the scalar field and the Ricci scalar $R$ together with the $R^2$ term. We separately analyzed the effect of such terms on the asymptotic bounce and tested for Conditions 1. and 2. for each theory.  Our analysis shows that:
\begin{itemize}
    \item In the presence of Einstein-Hilbert gravity, a bounce is allowed when a non-minimal coupling is included, while it is forbidden with a squared Ricci term;
    \item when the gravitational sector is scale-invariant (no Einstein-Hilbert term) a bounce is allowed when a non-minimal coupling is included if the field acquires a non-vanishing false vacuum value, which is larger than the width of the potential barrier. However, it is still forbidden when one adds a $R^{2}$ term. 
\end{itemize}

We believe that our method can be extended to include more general scenarios that are certainly worth exploring. In fact, finding the asymptotic bounce for massive theories would allow us to extend the flat spacetime result of \cite{Affleck:1980mp} to gravitational settings. This might have important implications on Higgs decay \cite{Isidori:2001bm,Degrassi:2012ry,Buttazzo:2013uya,Branchina:2014usa,Espinosa:2015kwx,Blum:2015rpa,Andreassen:2017rzq,Andreassen:2016cvx,Cabibbo:1979ay,Isidori:2007vm,Bezrukov:2012sa,Branchina:2013jra,Rajantie:2016hkj,Salvio:2016mvj,Czerwinska:2016fky,Bentivegna:2017qry,Branchina:2018xdh,Park:2018bkn,Branchina:2019tyy,Burda:2015yfa,Burda:2015isa,Gregory:2016xix,Gorbunov:2017fhq}, in which the mass term is usually neglected. Moreover,  vacuum decay from de Sitter spacetime is particularly important to test for Conditions 1. and 2 in the early and current Universe. Regarding the latter,  we should mention that existing calculations regarding Higgs decay with a non-minimal coupling are carried out in the flat spacetime approximation in the false vacuum state (see for example \cite{Rajantie:2016hkj,Branchina:2019tyy}). Our method may provide new windows in which other decay channels (such as decay through the Hawking-Moss instantons) dominate.

\subsection*{Acknowledgments}

 S.\ V.\ acknowledges the financial support of the Italian National Institute for Nuclear Physics (INFN) for her Doctoral studies. This work has been partially performed using the software Mathematica.
 \appendix
 \section{Coefficients $f_n$}
 \label{sec:appendix}
 Here, coefficients $f_n$ of Eq.\ \eqref{eq:higherorder.} are computed. Radial derivatives of arbitrary order are denoted by an index $(n)$, while radial derivatives of first and second order are denoted by one dot or two dots respectively. Derivatives of the potential with respect to the scalar field of  order $i$ are indicated as  $\dfrac{d^iV}{d\phi^i}$. As $f_n$s are computed at $\rho^*$ all quantities are implicitly evaluated at the turning point (recall that $\dot{\phi}_*=0$).
 Using Eq.\ \eqref{eq:eom} one gets
 \begin{equation}
 \label{eq:app_exp_eom1}
    \dfrac{d^iV}{d\phi^i}^{(n+1)}=\left(  \dfrac{d^{i+1}V}{d\phi^{i+1}}\dot{\phi}\right)^{(n)}\,,\qquad \phi^{(n)}=\left(  \dfrac{d^2V}{d\phi^2}\dot{\phi}\right)^{(n-3)}+\sum_{i=2}^{n-1}B_i \dfrac{\phi^{(i)}}{\rho^{n-i}}
 \end{equation}
 where $B_i$s are numerical factors, whose value is not relevant for the following discussion.
 
 Using the first equation in Eq.
 \eqref{eq:app_exp_eom1}, the (n+1)-th derivative of $V'$ may be written as
 \begin{equation}
 \begin{split}
 &  \dfrac{d V}{d\phi}^{(n+1)}=\\&= \dfrac{d^2V}{d\phi^2}\phi^{(n+1)}+\dots+   \left(\dfrac{d^2V}{d\phi^2}\right)^{(n-1)}\ddot{\phi}=\ddot{\phi}\left( \left(\dfrac{d^3V}{d\phi^3}\right)^{(n-3)}\ddot{\phi}+\dots+ \dfrac{d^3V}{d\phi^3}\phi^{(n-1)}\right)+\\&+\phi^{(3)}\left( \left(\dfrac{d^3V}{d\phi^3}\right)^{(n-4)}\ddot{\phi}+\dots+ \dfrac{d^3V}{d\phi^3}\phi^{(n-2)}\right)+\dots+\dfrac{d^2V}{d\phi^2}\phi^{(n+1)}
  \end{split}
  \end{equation}
  which can be further expanded using again Eq. \eqref{eq:app_exp_eom1}. Then one gets
 \begin{multline}
 \label{eq:appendix_exp}
     \dfrac{dV}{d\phi}^{(n+1)}= \dfrac{d^2V}{d\phi^2}\phi^{(n+1)}+\dfrac{d^3V}{d\phi^3}(\ddot{\phi}\,\phi^{(n-1)}+\phi^{(3)} \phi^{(n-2)}+\dots+\phi^{(n+1)/2}\phi^{(n+1)/2})+\\+\dfrac{d^4V}{d\phi^4}(\ddot{\phi}^2\phi^{(n-3)}+\phi^{(3)}\ddot{\phi} \phi^{(n-4)}+\dots+\phi^{(n+1)/3}\phi^{(n+1)/3}\phi^{(n+1)/3})+\dots.
 \end{multline}
Each $\dfrac{d^iV}{d\phi^i}$ in Eq.
\eqref{eq:appendix_exp} is multiplied by $i-1$ terms that are derivatives of $\ddot{\phi}$. They are of order $n+5-2i$ or lower, and thus these terms are non-vanishing only if $n+5-2i>1$. So, the highest-order derivative $\dfrac{d^{\bar{\imath}}V}{d\phi^{\bar{\imath}}}$ that appears in Eq. \eqref{eq:appendix_exp} is the one satisfying $n+5-2\,\bar{\imath}=3$ for even $n$ and $n+5-2\,\bar{\imath}=2$ for odd $n$.
 Expanding radial derivatives of $\ddot{\phi}$ in Eq.\eqref{eq:appendix_exp} using Eq.s\eqref{eq:app_exp_eom1}, $V'^{(n+1)}$ may be expressed in terms of derivatives of the potential with respect to the scalar field, $\ddot{\phi}=V'_*$ and $\rho^*$ only. From Eq.\eqref{eq:appendix_exp} one finds that the highest-order derivative (the $\bar{\imath}$-th term) is multiplied only by radial derivatives of the scalar field of order $2$ or $3$ and thus it contributes as
 \begin{equation}
     \dfrac{d^{\bar{\imath}}V}{d\phi^{\bar{\imath}}}\dfrac{\ddot{\phi}^{\bar{\imath}-1}}{\rho}\quad\text{even $n$,}\qquad  \dfrac{d^{\bar{i}}V}{d\phi^{\bar{\imath}}}\ddot{\phi}^{\bar{\imath}-1}\quad\text{odd $n$}
 \end{equation}
 to $f_n$.\\
 The second-highest derivative $\bar{\imath}-1$ is multiplied by radial derivatives of the scalar field of order $2,\,3,\,4,\,5$. Using Eq.\eqref{eq:app_exp_eom1}, derivatives of order $4$ and $5$ may be expressed in terms of lower derivatives. As can be seen from Eq.\eqref{eq:app_exp_eom1}, this results in an additional $V''$ contribution (numerical coefficients are omitted for simplicity)
 \begin{equation}
     \dfrac{d^{\bar{\imath}-1}V}{d\phi^{\bar{\imath}-1}}\dfrac{\ddot{\phi}^{\bar{\imath}-2}}{\rho^3}\left(1+A_1\dfrac{d^2V}{d\phi^2} \rho^2\right)\quad\text{even $n$,}\end{equation}
     \begin{equation*}\dfrac{d^{\bar{\imath}-1}V}{d\phi^{\bar{\imath}-1}}\dfrac{\ddot{\phi}^{\bar{\imath}-1}}{\rho^2}\left(1+A_2\dfrac{d^2V}{d\phi^2} \rho^2\right)\quad\text{odd $n$}.
 \end{equation*}
  The third-highest derivative $\bar{\imath}-2$ is multiplied by radial derivatives of the scalar field of order $2,\,3,\,4,\,5,\,6,\,7$. Using Eq.\eqref{eq:app_exp_eom1}, to express  derivatives of order $4,\,5,\,6,\,7$ in terms of lower order ones, one finds additional contributions with respect to the previous case
 \begin{equation}
     \dfrac{d^{\bar{\imath}-2}V}{d\phi^{\bar{\imath}-2}}\dfrac{\ddot{\phi}^{\bar{\imath}-3}}{\rho^5}\left(1+A_3\dfrac{d^2V}{d\phi^2} \rho^2+A_4\left(\dfrac{d^2V}{d\phi^2} \rho^2\right)^2+A_5\dfrac{d^3V}{d\phi^3} \rho^4 \ddot{\phi}\right)\quad\text{even $n$,}\end{equation}\begin{equation*} \dfrac{d^{\bar{\imath}-2}V}{d\phi^{\bar{\imath}-2}}\dfrac{\ddot{\phi}^{\bar{\imath}-3}}{\rho^4}\left(1+A_6\dfrac{d^2V}{d\phi^2} \rho^2+A_7\left(\dfrac{d^2V}{d\phi^2} \rho^2\right)^2+A_8\dfrac{d^3V}{d\phi^3} \rho^4 \ddot{\phi}\right)\quad\text{odd $n$}.
 \end{equation*}
In general, the $\bar{\imath}-i$th term has contributions from terms in Eq.\eqref{eq:appendix_exp} that are multiplied with a radial derivative of the scalar field of order $n-2i+3$ or higher. In this way, the dependence of the $\bar{\imath}-i$th term on $\ddot{\phi}$ and $\dfrac{d^jV}{d\phi^j}$ can be fully determined. The dependence on $\rho$ can be fixed by dimensional consistency. In particular, in each $\bar{\imath}-i$-th term, these contributions appear always in the combination
 $\dfrac{d^jV}{d\phi^j} \ddot{\phi}^{j-2}\rho^{2j-2}$  with $j\geq 2$.\\ Consider 
 \bea
      \lim_{t^*\rightarrow+\infty} \ddot{\phi}_* \rho^{4*}=0.
      \eea
 If all derivatives of the potential in the scalar field are finite for $\phi\rightarrow 0$, then \begin{equation}\label{eq:approx2}
     \dfrac{d^jV}{d\phi^j} \ddot{\phi_*}^{j-2} \rho^{*2j-2}\ll1\qquad \text{for $j>2$ and large $\rho^*$.}
 \end{equation}
 If also $\left(\dfrac{d^2V}{d\phi^2}\right)_*\rho^{*2}\ll1$, then
 \begin{equation}
    \left( \dfrac{dV}{d\phi}^{(n+1)}\right)_*\approx\sum_{i=0}^{i=\bar{\imath}-2}\tilde{A}_i\left(\dfrac{dV^{\bar{\imath}-i}}{d\phi^{\bar{\imath}-i}}\right)_* \dfrac{\ddot{\phi}_*^{\bar{\imath}-i-1}}{\rho^{*2i+1}}\quad\text{even $n$,}
 \end{equation}
 \begin{equation*}
    \left( \dfrac{dV}{d\phi}^{(n+1)}\right)_*\approx\sum_{i=0}^{i=\bar{\imath}-2}\tilde{A}_i\left(\dfrac{dV^{\bar{\imath}-i}}{d\phi^{\bar{\imath}-i}}\right)_* \dfrac{\ddot{\phi}_*^{\bar{\imath}-i-1}}{\rho^{*2i}}\quad\text{odd $n$,}
 \end{equation*}
 and the sum in Eq.\eqref{eq:higherorder.} is negligible with respect to $V'_*$ for large $\rho^*$ if Eq.\eqref{eq:approx2} holds.

\section{$S_C$ has a saddle point on the bounce}
\label{sec:appendixc}
$S_C$, as defined in Eq.\eqref{eq:sec1}, has a saddle point on the bounce, which can be turned into a minimum or maximum by slightly changing $S_{C,2}$. One has
 \begin{equation}\label{eq:dsdc}
 \dfrac{dS_C}{dC}=2 \pi^2\,\bar{t}^3 \left(V(\phi(\bar{t}))-V\left(\dfrac{C}{2 \bar{t}^2}\right)\right)+\pi^2\dfrac{C}{\bar{t}^2}+\pi^2\int_{\bar{t}}^{+\infty} \dfrac{dV}{d\phi}\left(\dfrac{C}{2t^2}\right) t \,dt+B_{\phi}+B_{g}.
 \end{equation}
 Here $B_{\phi}$, $B_g$ are boundary terms for the scalar field and for gravity, that appear by using the equations of motion in the variation of the first term of Eq.\eqref{eq:sec1}. There is one for the scalar field \begin{equation}
     B_{\phi}=\bar{t}^3\dot{\phi}(\bar{t})\,\dfrac{\delta \phi}{\delta C}(\bar{t})=-\dfrac{\pi^2 C}{\bar{t}^2},
 \end{equation}
 while the gravitational one can be computed from the Hawking-Gibbons-York boundary term \cite{York:1972sj,Gibbons:1976ue} evaluated at $t=\bar{t}$
 \begin{equation}
 \delta S_{GHY}=\oint_{\partial V}d^3x \epsilon \sqrt{|h|}n_{\alpha}V^{\alpha}   
 \end{equation} 
 with
  \begin{equation}
 V^{\alpha}=g^{\mu\nu}\delta\Gamma^{\alpha}_{\mu\nu}+g^{\alpha\mu}\delta\Gamma^{\nu}_{\mu\nu}  
 \end{equation}
 $$\delta \Gamma^{\alpha}_{\beta\gamma}=\dfrac{1}{2}g^{\alpha\mu}(\partial_{\beta}\delta g_{\gamma\mu}+\partial_{\gamma}\delta g_{\beta\mu}-\partial_{\mu}\delta g_{\gamma\beta})+\dfrac{1}{2}\delta g^{\alpha\mu}(\partial_{\beta} g_{\gamma\mu}+\partial_{\gamma} g_{\beta\mu}-\partial_{\mu} g_{\gamma\beta}) $$
and $\delta g_{\alpha\beta}$ is the variation of the metric, that has inverse $\delta g^{\alpha \beta}=-g^{\mu\alpha} g^{\nu\beta}\delta g_{\mu\nu}$.  Moreover, $n_{\alpha}$ is the unit normal to $\partial V$ and $h$ is the determinant of $h_{\alpha\beta}$, the induced metric on the boundary. $\epsilon$ is $+1$ if $\partial V$ is timelike, $-1$ if it is spacelike. Choosing a timelike future-oriented one-form $n_{\alpha}=(1,0,0,0)$ one gets
 \begin{equation}
 \label{eq:bt2}
     B_{g}=\dfrac{\pi}{16}\bar{t}^3 \left(2 g^{\alpha\beta}\delta \dot{g}_{\alpha\beta}+\delta g^{\alpha\beta}\dot{g}_{\alpha\beta}\right)=0
 \end{equation}
 
Thus the first term in Eq.\eqref{eq:dsdc} dominates and it gives
 \begin{equation}
     \dfrac{dS_C}{dC}\approx 0\qquad
     \dfrac{d^2S_C}{dC^2}\approx 0.
 \end{equation} One can turn the saddle point into a minimum or a maximum in the case of a single-scalar field with Einstein-Hilbert gravity and a non-minimal coupling. 
 To do that, $S_C$ may be redefined as \begin{equation}
 \label{eq:sec}
     S_C\equiv S_{C,1}+S_{C,2}
\end{equation}
where
\bea
     &&S_{C,1}\equiv2 \pi^2 \int_0^{\bar{t}} \bar{\rho}^3\left (\dfrac{\dot{\bar{\phi}}^2}{2}+V(\bar{\phi})-\dfrac{M_{\rm P}^2}{2}\bar{R}-\dfrac{\xi}{2}\bar{\phi}^2 \bar{R}\right)dt\\\non
     &&S_{C,2}\equiv2\pi^2\int_{\bar{t}}^\infty t^3 \left(\dfrac{C^2}{2 t^6}+V\left(\dfrac{C}{2 t^2}\right)\right)dt.
\eea 
Now
 \begin{equation}
 \label{eq:firstorder}
 \dfrac{dS_C}{dC}=-2 \pi^2 \dfrac{M_{\rm P}^2}{2} R\,\bar{t}^3\dfrac{d\bar{t}}{dC}-2 \pi^2 \dfrac{\xi}{2} R\,\phi^2 \bar{t}^3\dfrac{d\bar{t}}{dC}+\pi^2\int_{\bar{t}}^{+\infty} \dfrac{dV}{d\phi}\left(\dfrac{C}{2t^2}\right) t \,dt.
 \end{equation} To determine $\bar{t}(C)$ the bounce velocity is matched with the numerical estimate \begin{equation}
\label{eq:matching}
-\dfrac{C_0}{(\bar{t}+\delta)^3}=-\dfrac{C}{\bar{t}^3}
 \end{equation}
 where $\delta$ is a real number that satisfies $\rho(t)-t\approx\delta$ at large times on the bounce. It gives
 \begin{equation}
     C=C_0\left(1-\dfrac{3  \delta}{\bar{t}}\right)+\text{h.o.}
 \end{equation}
 where higher orders are suppressed at large $\bar{t}$ and for $\delta\neq 0$
. Thus 
\bea
 \dfrac{d\bar{t}}{dC}=\dfrac{\bar{t}^2}{C_0 3 \delta}
 \eea
 to lowest order.
 Eq.\eqref{eq:firstorder} then gives
 \begin{equation}
     \dfrac{dS_C}{dC}\approx-\dfrac{\pi}{8} R(\bar{t})\,\bar{t}^3\dfrac{d\bar{t}}{dC}\approx-\dfrac{C_0 \pi^2}{3 \delta\bar{t}}=-\dfrac{C_0(C_0-C)\pi^2}{9\delta^2}(1+6\xi)\qquad
     \dfrac{d^2S_C}{dC^2}\approx \dfrac{\pi^2}{9 \delta^2}(1+6\xi)
 \end{equation}
 $S_C$ has a minimum on the bounce for $\xi>-\dfrac{1}{6}$ and a maximum otherwise.
\section{Bounce action from a perturbative expansion}
\label{sec:appendix:pertexp}
Some concerns have been raised  in the literature  \cite{Rajantie:2016hkj,Branchina:2016bws} as regards the use of a perturbative expansion to determine the bounce action of a scalar field theory with an Einstein-Hilbert term as done in \cite{Isidori:2007vm}. To do that, one expands the scalar field and the scale factor around the bounce in flat space as
\bea
\phi(t)=\phi_0(t)+\kappa \phi_1(t)\\
\rho(t)=t+\kappa \rho_1(t).
\eea 
Using the equations of motion, the action is expanded in the same way and one finds that the on-shell action is determined by
\bea
S=S_0+\dfrac{A}{M_P^2 \mathcal{R}^2}\qquad \text{for}\qquad M_P^2 \mathcal{R}^2\gg 1
\eea
with $S_0$ the flat space on-shell action, $A$ a real constant and $\mathcal{R}$ the bounce radius, $\phi_0=2 \mathcal{R}$. Then, it is minimized with respect to $R.$ Actually, it has been shown \cite{Rajantie:2016hkj} that $\phi_1$ in a scalar field theory with negative quartic potential does not satisfy the proper boundary conditions and thus the perturbative expansion is unreliable. There might be, however, an alternative interpretation of the calculation that makes the final result justified. Consider two theories that have a bounce, and one of them is the vanishing coupling limit of the other (in this case it is the single scalar field theory arising from the $M_P \mathcal{R}\rightarrow+\infty$ limit of the same theory with Einstein-Hilbert gravity). By continuity, the action, as the coupling is turned on, changes by a small amount, for sufficiently small values of the coupling. Moreover, if the gravitational backreaction is small, one might choose an \emph{off-shell} profile with the same shape as $\phi_0(t)$, as a function of an arbitrary parameter $\mathcal{R}$, and use it to determine a correction for $\rho_1(t)$ and thus $S_E$, which is to be minimized as a function of $\mathcal{R}$. If the action has a minimum, the off-shell profile is, to first order in $O(M_P^{-2})$, the bounce one of the full theory. In this way, one avoids to use the perturbation equation for the scalar field, and uses the flat space solution only as a field profile to keep the action functional finite.\\
To keep the approximation under control is instead trickier. While the natural choice might be requiring $M_P\mathcal{R}\gg 1$ one can see that a transformation that leaves the equations of motion Eq.\eqref{eq:eom1EH}- \eqref{eq:eom2EH}unchanged
 \bea
 V(\phi)\rightarrow \alpha V(\phi)\qquad t\rightarrow t \,\alpha^{-1/2}\qquad\rho\rightarrow \rho    \,\alpha^{-1/2}\qquad \text{with real $\alpha$}
 \eea
 changes the action as $\alpha S_E\rightarrow S_E$. Changing $\alpha,$  $\mathcal{R}$ can be arbitrarily small, increasing $\lambda$ accordingly.  In this case, the relevant parameter is thus $M_P\mathcal{R}\sqrt{\lambda}$. The detailed generalization to theories with multiple couplings $\lambda_i$ seems to be more involved, despite one may guess that \bea
M_P\mathcal{R}\sqrt{\lambda}_i\gg 1\eea
keeps the gravitational backreaction small.

\end{document}